\documentclass{appolb}
\usepackage{graphicx}

\usepackage{a4p}
\usepackage{rotating}

\usepackage{subfigure}
\usepackage{longtable}
\usepackage{multirow}
\usepackage{textcomp}
\usepackage{authblk,colortbl}
\usepackage{units}
\usepackage{amsmath,amssymb,amsfonts} 
\usepackage{mathrsfs}
\usepackage{color} 
\usepackage{authblk}
\usepackage{hyperref} 

\def\beq{\begin{equation}} 
\def\eeq{\end{equation}}
\def\bea{\begin{eqnarray}}
\def\eea{\end{eqnarray}}
\def\eqref#1{eq.~(\ref{eq:#1})}

\def\nn{\nonumber}   

\begin{document}

\title{Spin effects in the tau-lepton pair induced by anomalous magnetic and electric 
dipole moments\thanks{Presented at the XLVI International Conference of Theoretical Physics ``Matter to
the Deepest'', Katowice, Poland, 15-19 September, 2025.}
} 
\author{A.~Yu.~Korchin
\address{NSC Kharkiv Institute of Physics and Technology, 61108 Kharkiv, Ukraine } 
\address{V.~N.~Karazin Kharkiv National University, 61022 Kharkiv, Ukraine } 
\address{M.~Smoluchowski Institute of Physics, Jagiellonian University, 30-348 Krakow, Poland} 
}

\maketitle

\begin{abstract}
The possible anomalous New Physics contributions to magnetic and electric dipole moments of the $\tau$ lepton 
have brought renewed interest in studying $\tau$-pair production at energies of the LHC and future colliders. 
We discuss effects of electromagnetic and weak dipole moment contributions 
to the $\tau$-lepton polarization and $\tau \tau$ spin correlations in the  
$\gamma\gamma \to \tau^-\tau^+$  and $q \bar{q} \to \tau^- \tau^+$ processes.  
Such processes have been observed in $pp$ and PbPb collisions in the LHC experiments.
Extensions of the Standard Model amplitudes for $\gamma\gamma \to \tau^-\tau^+$ and 
$q \bar{q} \to \tau^- \tau^+$ processes, which include dipole moments of the $\tau$ lepton, 
are implemented in the {\tt TauSpinner} Monte Carlo program. 
A few examples of signatures of $\tau \tau$ spin correlations and $\tau$-lepton dipole moments 
in observables are presented.
\end{abstract}


%

\section{Introduction}
\label{sec:intro}

Recent measurements of dipole moments of the $\tau$ lepton at the Belle experiment~\cite{Belle:2021ybo},
as well as observation of $\gamma\gamma \to \tau^-\tau^+$ production at the LHC~\cite{ATLAS:2022ryk,CMS:2022arf} have brought considerable interest in magnetic and electric dipole moments of the $\tau$ lepton.
Deviation from predictions of the Standard Model (SM) and measured values can carry information on the 
New Physics (NP) effects, which are expected to be significantly enhanced for the $\tau$ lepton 
compared to the muon. Various extensions of the SM, for example, new heavy 
particles in the loops, can be a source of NP contributions to dipole moments of the $\tau$ lepton,
as mentioned in~\cite{Bernreuther:1996dr, Eidelman:2007sb} and references therein. 

In our studies the anomalous magnetic and electric dipole moments of the $\tau$ lepton, 
or the corresponding form-factors in case of virtual photon and $Z$ boson, are introduced on top of 
simulations of $\gamma\gamma \to \tau \tau$ and $q \bar{q} \to \tau \tau$ processes 
in the SM including decays of the $\tau$ leptons. The important ingredient in this calculation, 
which complicates consideration, is account of the spin correlations in the $\tau$ pair. 
The Monte Carlo solutions are convenient and calculations are 
implemented in the {\tt TauSpinner} program~\cite{Czyczula:2012ny, Przedzinski:2018ett} 
for reweighting events with $\tau$ leptons produced in the $pp$ or PbPb collisions.

In the calculation of the hard quark-antiquark processes, the Improved Born Approximation (IBA) is applied, which 
includes electroweak (EW) corrections following the formalism outlined in 
Refs.~\cite{Bardin:1999yd, Richter-Was:2018lld}.

In this article, some aspects of the spin-correlation approach  
to the $\gamma\gamma \to \tau \tau$ and $q \bar{q} \to \tau \tau$ processes are reviewed. 
A few examples of impact of the $\tau$-lepton dipole moments on observables in 
$pp$ collisions are illustrated. Results presented here are 
based on Refs.~\cite{Banerjee:2023qjc, Korchin:2024appb, Korchin:2025vzx,Korchin:2025sui}.      
   

\section{Magnetic and electric dipole form-factors and moments}
\label{sec:dipole moments}

Magnetic and electric dipole moments of the $\tau$ lepton induce its interaction with external 
magnetic (${\cal \vec{B}}$) and electric (${\cal \vec{E}}$) fields
\beq 
H_{int} = -  \vec{\mu}_\tau \, {\cal \vec{B}} - \vec{d}_\tau \,  {\cal \vec{E}},
\label{eq:01}
\eeq
where in the rest frame of the lepton, vectors of dipole moments are proportional to the $\tau$-lepton 
spin $\vec{s}$:  
\beq 
\vec{\mu}_\tau  =  g_\tau \frac{e Q_\tau }{2m_\tau} \vec{s},    \quad \qquad  a_\tau = \frac{g_\tau}{2}-1,
\label{eq:02}
\eeq
\beq 
\vec{d}_\tau =  \eta_\tau \frac{e}{2 m_\tau} \vec{s}, \qquad  \quad   d_\tau 
= \frac{\eta_\tau}{2} \frac{e }{2m_\tau}. 
\label{eq:03}
\eeq
Here $g_\tau$ is the $g$-factor, $\eta_\tau$ is the dimensionless constant analogous to $g_\tau$, 
$a_\tau$ is the anomalous magnetic dipole moment, and $d_\tau $ is the electric dipole moment in units $e\,$cm.

At high energies one can use Lorentz-covariant form of the $\gamma \tau \tau$ 
vertex\footnote{The anapole form-factor $F_A(q^2)$ is not considered here.} 
 \beq \Gamma_\gamma^\mu 
= -ie Q_\tau \Bigl\{ F_1(q^2) \gamma^\mu + \frac{\sigma^{\mu \nu } q_\nu}{2 m_\tau} 
\bigl[ i \, F_2(q^2) + \gamma_5 \, F_3(q^2) \bigr]  \Bigr\}
\label{eq:Gamma_gamma} 
\eeq
in terms of the Dirac  ($F_1(q^2)$), Pauli ($F_2(q^2)$) and electric dipole ($F_3(q^2)$)  form-factors, 
which reduce to the dipole moments at the real-photon point at $q^2=0$
\beq  
F_2(0) =  a_\tau,  \qquad F_3(0) = \frac{2m_\tau}{e Q_\tau} d_\tau,  \qquad F_1(0)= 1.  
\label{eq:FF}
\eeq
In the following, we use the notation $A(q^2) \equiv F_2(q^2)$ and $B(q^2) \equiv F_3(q^2)$.  

At high energies of the LHC and future colliders, the $Z$-boson interaction 
with the $\tau$ lepton (in general, with any fermion) becomes important, and   
the $Z \tau \tau$ vertex including the SM contribution and dipole moment terms can be written as    
\begin{equation}
\Gamma^\mu_Z  =  { -i} \frac{g_Z}{2} \, \Big\{ \gamma^\mu (g_V^\tau - \gamma_5 g_A^\tau ) 
+ \frac{\sigma^{\mu \nu} q_\nu}{2 m_\tau}  \big[ i  X(q^2)  +  \gamma_5  Y(q^2)   \big]  \Big\}.
\label{eq:Gamma_Z}  
\end{equation} 
Here $g_Z =e/(s_W c_W) $ is the constant of the neutral-current interaction,   
$s_W \equiv \sin \theta_W$,  $c_W \equiv \cos \theta_W$, where  $\theta_W$ is the weak mixing angle,   
and $g_V^\tau$,  $g_A^\tau$ are the vector and axial-vector couplings. 

In Eq.~(\ref{eq:Gamma_Z}), $X(q^2)$ is the weak anomalous magnetic form-factor ($CP$ conserving), 
and $Y(q^2)$ is the weak electric form-factor ($CP$ violating),  which on the $Z$-boson mass shell 
at $q^2=M_Z^2$ are related to the weak dipole moments as
\beq 
X(M_Z^2) = a_\tau^{(w)}  (2 s_W c_W),  \qquad Y(M_Z^2) =  d_\tau^{(w)}  (2 s_W c_W).
\label{eq:X_Y}
\eeq 
The weak dipole moments $a_\tau^{(w)}$ and $d_\tau^{(w)}$ are defined by ALEPH 
collaboration in Ref.~\cite{ALEPH:2002kbp}.   

In general, one has to separate the SM radiative corrections to the dipole moments, or form-factors, 
from NP contributions.    
For the real photons, the SM anomalous magnetic dipole moment was calculated in~\cite{Eidelman:2007sb} 
with the result $a_{\tau, \, {SM}}= (117721 \pm 5) \times 10^{-8}$, and more 
recently in~\cite{Keshavarzi:2019abf},  where $a_{\tau, \, SM} =  (117717.1 \pm 3.9) \times 10^{-8}$. 
For the finite $q^2$, one can use the QED contribution to $A(q^2)$ in the first order 
in $\alpha = e^2/(4 \pi)$~\cite{Berestetskii:1982qgu}. 
 
The electric dipole moment is equal to $d_{\tau, \, SM} =  7.32 \times 10^{-38}\; e\,$cm, which was 
calculated~\cite{Yamaguchi:2020dsy} in the SM including long-distance contributions.    
The SM prediction~\cite{Bernabeu:1994wh} for the weak anomalous magnetic moment is
$a_{\tau, \, SM}^{(w)} = - (2.10 + i \, 0.61) \times 10^{-6}$.   

In our calculation, we neglect minor contributions from the SM to the electric form-factor $B(q^2)$, 
as well as to the weak form-factors $X(q^2)$ and $Y(q^2)$. These form-factors 
can be viewed as originating mainly from NP. As for the magnetic form-factor $A(q^2)$, the 
SM term is added via $A(q^2) = A(q^2)_{SM} + A(q^2)_{NP}$. To simplify further notation, we will not 
distinguish between form-factors and corresponding dipole moments for the NP contribution, and 
will denote them by $A_{NP}, \, B_{NP}=B, \, X_{NP}=X$ and $Y_{NP}=Y$.    

We should mention important relations between the NP contributions to these 
dipole moments and Wilson coefficients of the SM Effective Field Theory (SMEFT).
The SMEFT Lagrangian is written as~\cite{Buchmuller:1985jz, Aguilar-Saavedra:2008nuh, Grzadkowski:2010es}
\beq
 {\cal L}_{SMEFT} = {\cal L}^{(4)}_{SM}+ 
\Bigl( \frac{1}{\Lambda} \sum_k C_k^{(5)} Q^{(5)}_k + 
\frac{1}{\Lambda^2} \sum_k C_k^{(6)} Q^{(6)}_k + {\cal O}\Bigl( \frac{1}{\Lambda^3} \Bigr) 
+ {\rm H.c.} \Bigr),   
\label{eq:SMEFT}
\eeq 
where ${\cal L}^{(4)}_{SM}$ is the SM Lagrangian and $\Lambda$ is the scale of NP. 
 
Relevant for the dipole moments terms are the dimension-6 operators 
\beq  
Q_{\tau B}^{(6)} =  \bigl(\bar{L}_L  \sigma^{\mu \nu}  \tau_R \bigr) \, \varphi \,  B_{\mu \nu}, \qquad 
\quad Q_{\tau W}^{(6)} =   \bigl( \bar{L}_L  \sigma^{\mu \nu}  \tau_R \bigr)  
\vec{\sigma} \, \varphi  \, \overrightarrow{W}_{\mu \nu} 
\label{eq:QB-QW}
\eeq 
which enter Eq.~(\ref{eq:SMEFT}) with the corresponding dimensionless Wilson coefficients 
$C^{(6)}_{\tau B}$ and $C^{(6)}_{\tau W}$. In Eq.~(\ref{eq:QB-QW}), $\varphi$ is the scalar doublet, $B_{\mu \nu}$ and $\overrightarrow{W}_{\mu \nu}$ are the strength tensors for the $U(1)_Y$  and $SU(2)_L$ gauge fields $B_\mu$ and $\overrightarrow{W}_\mu$, $\tau_R$ is right-handed chiral 
spinor and $\bar{L}_L = (\bar{\nu}_{\tau } \, \bar{\tau})_L$ contains left-handed chiral adjont spinors.        

Straightforward calculation leads to the relations between the NP contributions to the   
dipole moments and Wilson coefficients 
\bea
\label{eq:A_B_X_Y}
&& A_{NP}  = \lambda \, {\rm Re} ( {D_{\gamma \tau \tau}}),  \qquad \qquad \qquad  \; \; \; \;
B_{NP} =  - \lambda \, {\rm Im} ({D_{\gamma \tau \tau}} ),  \\
&& X_{NP}  =   \lambda \,  {\rm Re} ({D_{Z  \tau \tau}}) \, (2 s_W c_W),  \qquad \quad 
Y_{NP} =   - \lambda \,  {\rm Im} ({D_{Z  \tau \tau}}) \, (2 s_W c_W),   \nn
\eea		
where 
\beq 
\lambda = \frac{\sqrt{2} \,  v }{\Lambda^2} \, \frac{2 m_\tau}{e},  
\label{eq:lambda}
\eeq 
\bea  
{D_{\gamma \tau \tau }} =  c_W  C^{(6)}_{\tau B} - s_W  C^{(6)}_{\tau W}, \qquad \qquad
 {D_{Z \tau \tau}} =  s_W  C^{(6)}_{\tau B}  +  c_W C^{(6)}_{\tau W}   
\eea
and $v = (\sqrt{2} G_F)^{-1/2} \approx 246$ GeV is the vacuum expectation value of the scalar field. 

Once dipole moments are measured, the coefficients
$C^{(6)}_{\tau B}/\Lambda^2$ and $C^{(6)}_{\tau W}/\Lambda^2$ can be determined  
which will indirectly constrain an underlying high-energy theory. 		


\section{\texorpdfstring{Spin correlations in the $\tau$ pair}{}}
\label{sec:spin correlations}

Many details on inclusion of spins of the final $\tau$ leptons in the processes  
$e^- e^+ \to \tau^- \tau^+$, \ $q \bar{q} \to \tau^- \tau^+$ and $\gamma \gamma \to \tau^- \tau^+$ 
are contained in Refs.~\cite{Banerjee:2023qjc, Korchin:2025vzx, Banerjee:2022sgf}. Below a few  
aspects of this approach are recalled.    

The process $a \, b \to \tau^+\tau^-$  with $(a \, b)$ = $(e^-  e^+)$,  $(q \, \bar{q})$ or 
$(\gamma \gamma)$ 
\begin{equation}
a (k_1) + b (k_2) \to \tau^- (p_-) + \tau^+ (p_+)
\label{eq:001}
\end{equation}
is described in the center-of-mass frame, where components of four-momenta are chosen as  
\begin{eqnarray}
&& p_- =(E, \vec{p}), \qquad \;  p_+ =(E, \, -\vec{p}), \qquad  \vec{p} = (0, \, 0, \, p),   \\ 
&& k_1 = (E, \, \vec{k}) , \qquad k_2  = (E, \, -\vec{k}),  \qquad 
\vec{k} =  (E \, \sin \theta, \, 0, \, E \, \cos \theta ),  \nonumber
\label{eq:002}
\end{eqnarray}
so that $\hat{3}$ axis is along vector $\vec{p}$, the reaction plane is spanned on 
$\hat{1}$ and $\hat{3}$ axes defined by the momenta $\vec{p}$ and $\vec{k}$, 
and $\hat{2}$ axis is along $\vec{p} \times \vec{k}$. Also,
$E=\tfrac{1}{2} \sqrt{s}$ is the energy, $p = \beta E$ is the momentum, 
and $\beta = (1-4 \tfrac{m_\tau^2}{s})^{1/2}$ is the velocity of the $\tau$ lepton.
  
After squaring the matrix element and averaging result over the polarization states 
of the initial fermions or photons, one obtains
\begin{equation}
|{\cal M}|^2 =    \sum_{i, j=1}^4 \, R_{i j} \, s^-_i  s^+_j  
= R_{44} \Bigl(1 + \sum_{i=1}^3 \, r_{i 4} \,  s^-_i +  \sum_{j=1}^3 \, r_{4 j} \, s^+_j + 
\sum_{i, j = 1}^3 \, r_{i j} \, s^-_i  s^+_j \Bigr).  
\label{eq:010}
\end{equation}
Here  $ \vec{s}^{\, +} =(s^+_1, \, s^+_2, \, s^+_3)$ and 
$\vec{s}^{\, -}=(s^-_1, \, s^-_2, \, s^-_3) $ are vectors of spins of the $\tau^+$ and $\tau^-$ 
leptons in their corresponding rest frames, and for convenience we add the 4th components 
equal to $1$, so that $s^\pm_i \equiv (\vec{s}^{\, \pm}, \, 1)$ for $i, j =1, 2, 3, 4$.  
 
The elements of the spin-correlation matrix $R_{ij}$ depend on the invariant mass of 
the $\tau$ pair, $m_{\tau \tau}$, and the scattering angle $\theta$. 
In Eq.~(\ref{eq:010}), the normalized elements of the matrix 
are introduced via $r_{i j } \equiv R_{ij}/R_{44}$ ($i, j = 1,2,3,4$). 
The spin-independent element $R_{44}$ determines the cross section 
for the unpolarized $\tau$ leptons.

Elements $R_{i j}$  were calculated in Refs.~\cite{Banerjee:2023qjc, Korchin:2025vzx, Banerjee:2022sgf}. 
For the $\gamma \gamma \to \tau \tau$ reaction, the tree-level $t$- and $u$-channel diagrams were 
taken into account including dipole moments up to the $4^{th}$ order~\cite{Korchin:2025vzx}.  
For the  $q \bar{q} \to \tau \tau$ processes,  the $s$-channel exchange 
of virtual $\gamma$ and $Z$ boson was included,  on top of which the EW corrections in the IBA were added~\cite{Banerjee:2023qjc, Korchin:2025sui}. The contribution from the dipole form-factors was accounted 
for in the lowest order.  

The elements $(r_{14}, r_{24}, r_{34})=\vec{{\cal P}}(\tau^-)$ 
and $(r_{41}, r_{42}, r_{43})=\vec{{\cal P}}(\tau^+)$ are components of polarization 
of the $\tau^-$ and $\tau^+$ leptons.  These elements vanish for the $\gamma \gamma \to \tau \tau$ 
reaction in a good approximation, while they are non-zero in the case of the $q \bar{q} \to \tau \tau$ process.
The elements $r_{ij}$ with $i, j=1,2,3$ form $3 \times 3$ matrix of $\tau^- \tau^+$ spin correlations.    

Note that matrix $R_{ij}$ has certain symmetry properties. In general, the following relations hold: 
\beq
R_{j i} =  R_{i j}\big|_ {A \to A, \, B \to -B, \, X \to X, \, Y \to -Y} \; \; \; \; {\rm for } \; \; \; i \ne j,
\label{eq:symmetries}
\eeq
and some non-diagonal elements vanish if the dipole moments (form-factors) are switched off. 
In particular, if electric dipole moment is absent, then for the $\gamma \gamma \to \tau \tau$ process        
$R_{12}=R_{21}=R_{23}=R_{32}=0$.
 

\section{Results and discussion}
\label{sec:results}
In this section, a few numerical results for the $\gamma \gamma \to \tau \tau$ 
and $q \bar{q} \to \tau \tau$ processes are presented. We will 
discuss some observables, which can be studied in the experimental analyses. 

Let us mention that the frame and sign convention of $R_{ij}$ defined above 
differ from the ones used in the {\tt TauSpinner} program,
when the matrix is contracted with $\tau$-lepton polarimetric vectors. 
There are reasons for that frame orientation differences used in {\tt TauSpinner}
and {\tt Tauola} decay library~\cite{Jadach:1993hs}, they are discussed in 
Ref.~\cite{Korchin:2025vzx}. The result of redefinition of the spin-correlation elements is   
\begin{eqnarray}
\label{eq:frames}
R_{tt}= \;\; R_{44}, & R_{tx}=     -R_{42}, & R_{ty}=     -R_{41},\;  R_{tz}=     -R_{43},  \\
R_{xt}=     -R_{24}, & R_{xx}= \;\; R_{22}, & R_{xy}= \;\; R_{21},\;  R_{xz}= \;\; R_{23}, \nonumber          \\
R_{yt}= \;\; R_{14}, & R_{yx}=     -R_{12}, & R_{yy}=     -R_{11},\;  R_{yz}=     -R_{13}, \nonumber \\
R_{zt}=     -R_{34}, & R_{zx}= \;\; R_{32}, & R_{zy}= \;\; R_{31},\;  R_{zz}= \;\; R_{33}. \nonumber 
\end{eqnarray}
With this transformation, $R_{ij} $ with $i, j=t, x, y, z$ are used  in the {\tt TauSpinner} event  reweighting algorithm for calculating weight implemented in the {\tt TauSpinner} code and for presenting numerical results.


\subsection{\texorpdfstring{Spin-correlation effects in $\gamma \gamma \to \tau \tau$ process}{}}
\label{subsec:gamma-gamma}

Results presented below are based on events generated with {\tt Pythia 8.3}~\cite{Bierlich:2022pfr}
using $pp$ scattering at 13 TeV,  {\tt PhotonCollision:gmgm2tautau} hard process, 
parametrization of parton distribution functions {\tt PDF:pSet = 13} and restricted to low invariant mass 
of the $\tau$ pair, $m_{\tau\tau} =$ 5-50 GeV and transverse momentum 
$p_T^{\tau\tau} >$ 5 GeV. This choice of phase-space roughly corresponds to the range
covered by the PbPb$\to$Pb($\gamma \gamma \to \tau \tau$)Pb
processes in peripheral PbPb collisions at the LHC. 

The $\tau$ decays were modeled with {\tt Tauola} decay library~\cite{Jadach:1993hs} (with no spin correlations between decaying $\tau^+ \tau^-$ leptons).  In total there are about $0.8 \times 10^6$ events for each decay mode generated. Then the spin correlations were added using weight calculated with 
{\tt TauSpinner} program,  both for the spin-correlation effects in the SM and in SM+NP models, 
and the cross-section normalization in SM+NP models.

We use as a reference the SM calculation with $A=B=0$, and assume for simplicity that
the QED radiative corrections are small enough and can be omitted when discussing the NP effects. 
Thus any non-zero values of $A, \, B$ indicate some NP.  

Below the SM predictions are compared with the six settings for the dipole moments 
\bea
\label{eq:settings}
&&  \; (i) \; \, A=0.002, \, 0005, \, 0.02 \;\, {\rm with} \; \,  B=0, \\
&&  \; (ii) \; B=0.002, \, 0005, \, 0.02 \;\, {\rm with} \; \,  A=0. \nn
\eea
This choice is arbitrary covering the plausible range. The 
smallest value $A=0.002$ roughly corresponds to $2 \, a_{\tau, \, SM}$.

The analysis shows that behavior of the matrix $r_{ij}$ is not trivial. In some cases, 
its elements are zero in the SM, while they obtain contribution from the dipole moments.  
In many cases the spin correlations in the SM are sizable and dominate over impact from the
dipole moments. This can give hints on how to optimize choice of the observables
and minimize background, and at the same time, underlying importance of spin correlations 
as a possible bias for the cross-sections.

\begin{figure}[hbt!]
  \begin{center}                               
{ \includegraphics[width=5.7cm,angle=0]{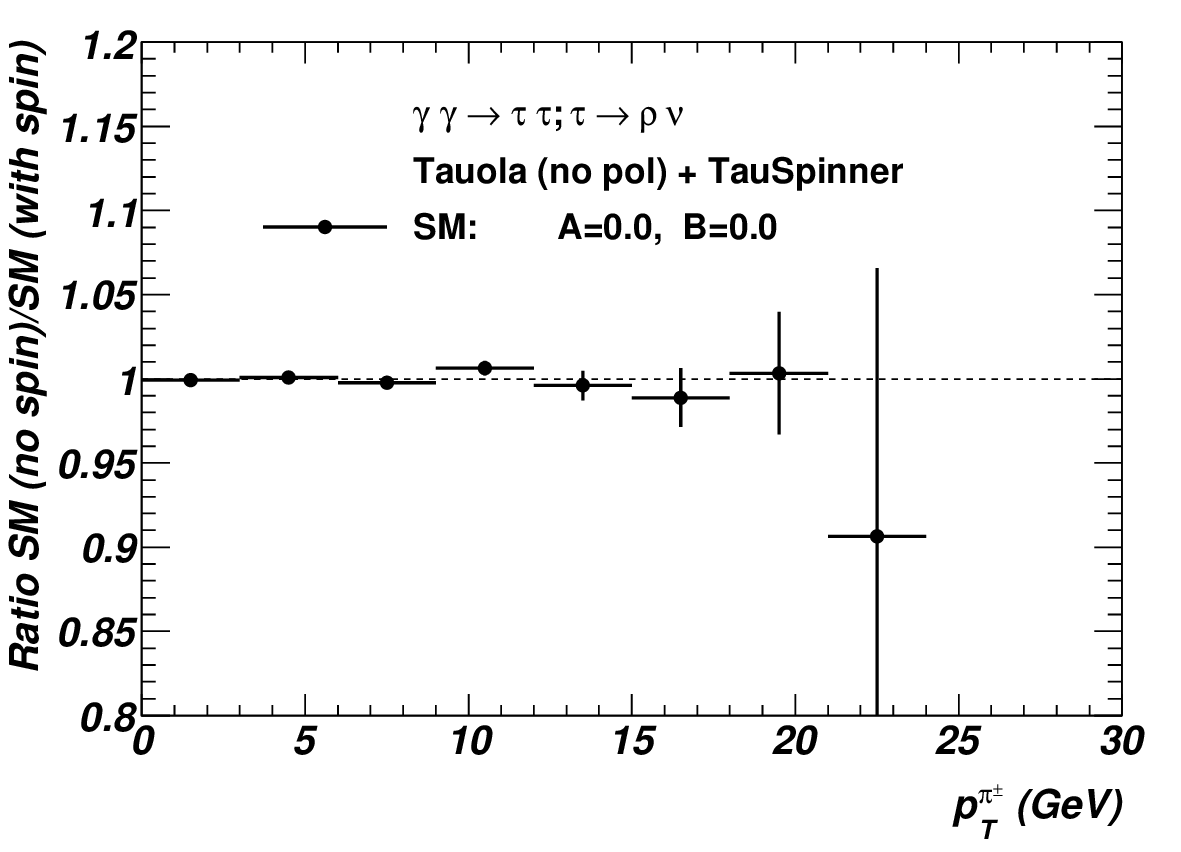}
   \includegraphics[width=5.7cm,angle=0]{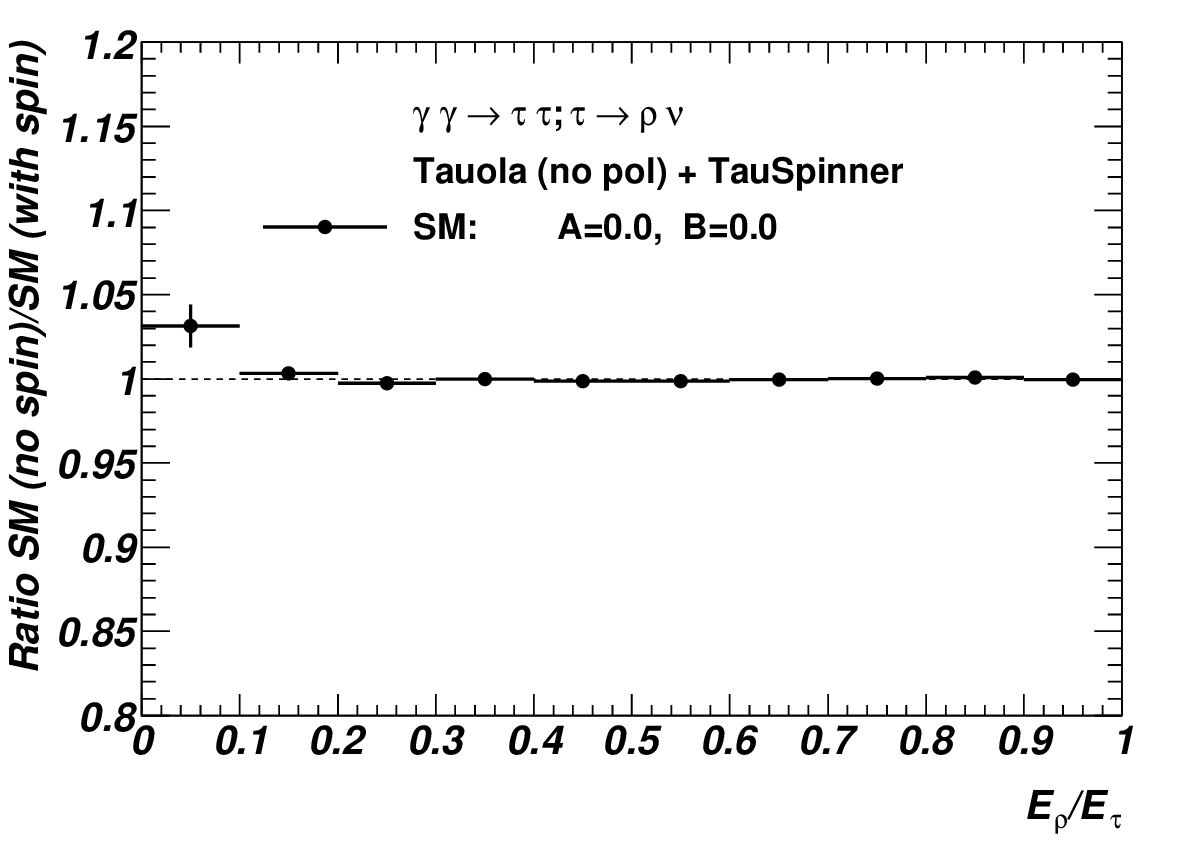} \\
   \includegraphics[width=5.7cm,angle=0]{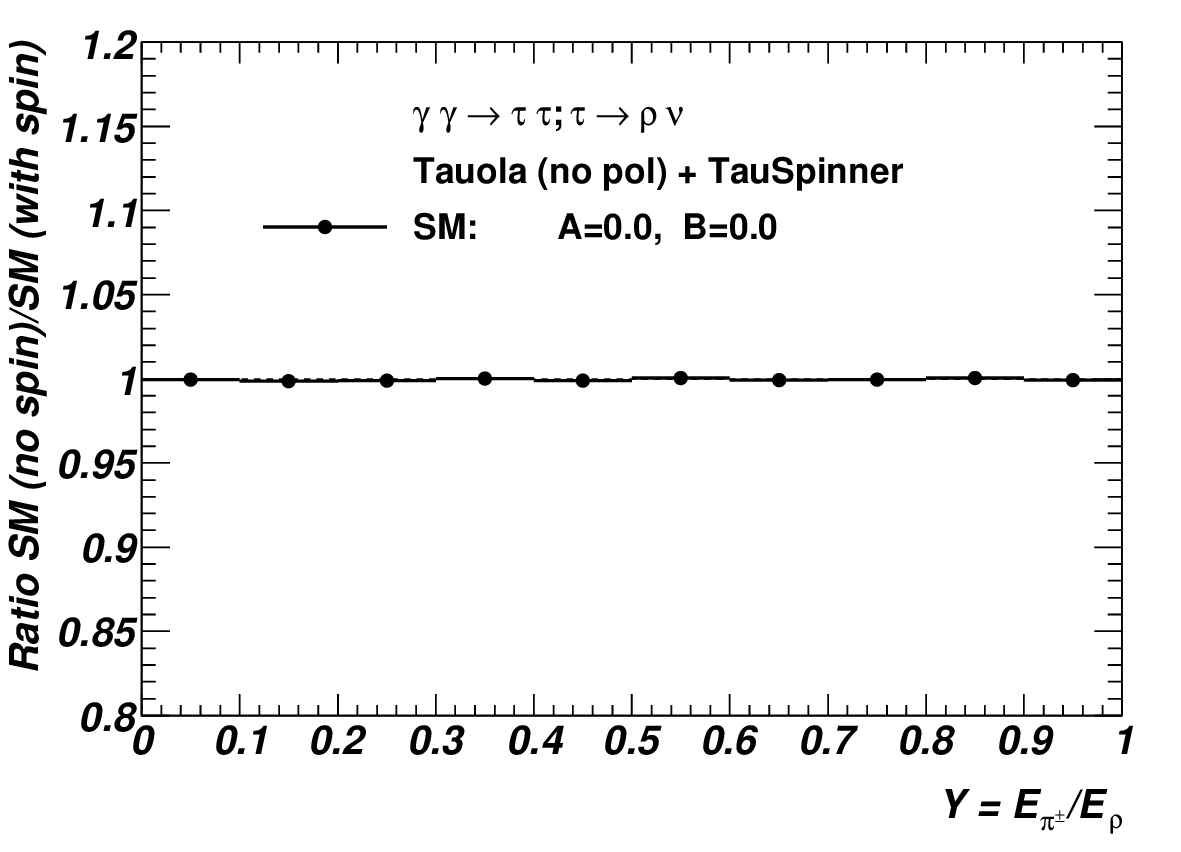}
   \includegraphics[width=5.7cm,angle=0]{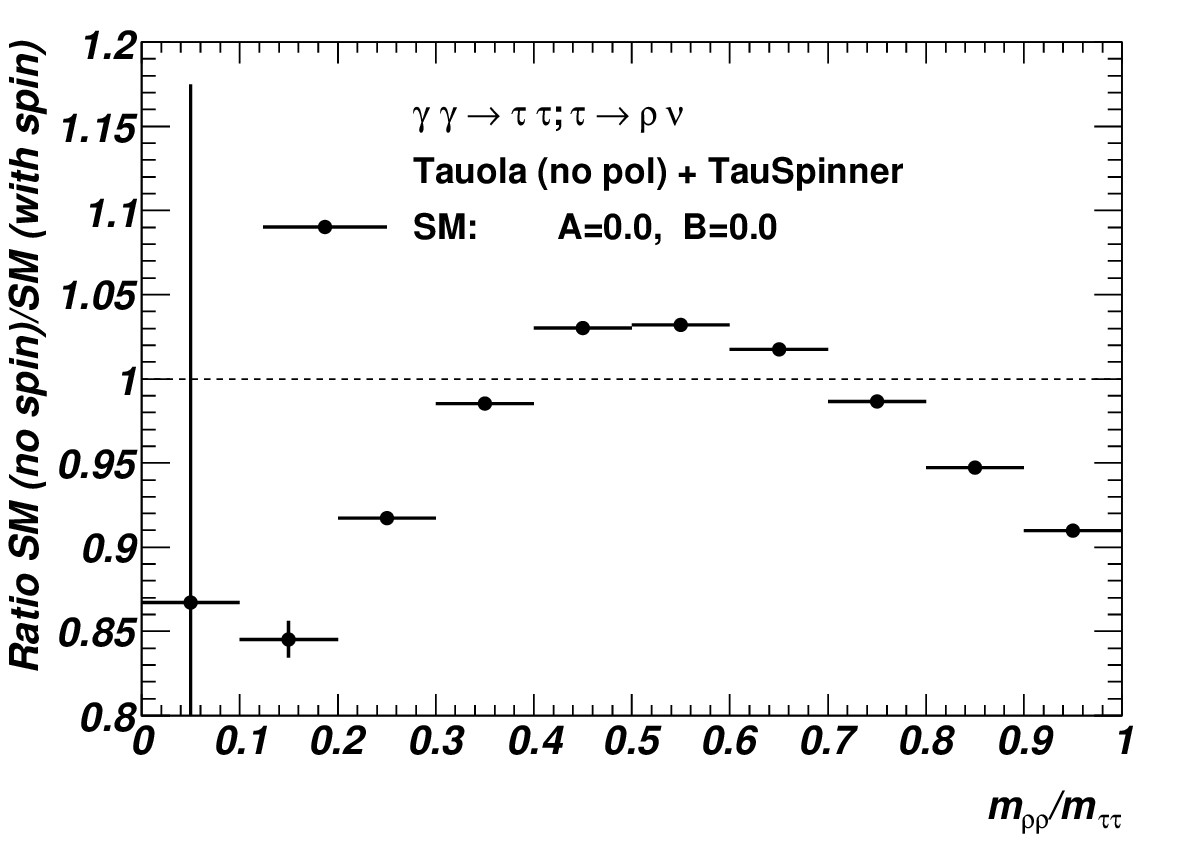}}
\end{center}
  \caption{Spin-correlation effects in $\gamma \gamma \to \tau \tau$ for $\tau$-decay channels:  
	$\tau^-\to \rho^- \nu_\tau \to \pi^- \pi^0 \nu_\tau$ and $\tau^+ \to \rho^+ \bar{\nu}_\tau \to 
	\pi^+ \pi^0 \bar{\nu}_\tau$. 
   The ratio (SM without spin correlations)/(SM with spin correlations) is shown.
	The figure is taken from Ref.~\cite{Korchin:2025vzx}.}
 \label{Fig:kinem_SM_rhorho} 
\end{figure}

Let us discuss distributions constructed from momenta of the $\tau$-decay products. 
Below example of the decay channels $\tau^- \to \rho^- \nu_\tau \to \pi^- \pi^0 \nu_\tau$
and $\tau^+ \to \rho^+ \bar{\nu}_\tau \to \pi^+ \pi^0 \bar{\nu}_\tau$ is considered. 

In Fig.~\ref{Fig:kinem_SM_rhorho}, the effect of spin correlations in the SM with $A=B=0$ is shown
on a few distributions:  \\
(i) transverse momenta of charged pions $p_T^{\pi}$, \\
(ii) ratios $E_{\rho}/E_{\tau}$ and $E_{\pi^\pm}/E_{\rho}$, \\
(iii) ratio of invariant mass of $\rho^+\rho^-$ system to $\tau^+\tau^-$ system, 
$m_{\rho\rho}/m_{\tau\tau}$.

The $p_T^{\pi}$,  $E_{\rho}/E_{\tau}$ and $E_{\pi^\pm}/E_{\rho}$ distributions are 
insensitive to the spin correlations. However, in the distribution of $m_{\rho \rho}/m_{\tau\tau}$, 
the effect is apparent. The change in the shape of the latter is   
10-15\% in a wide range around $m_{\rho\rho}/m_{\tau\tau} \approx 0.5$. 

\begin{figure}[hbt!]
  \begin{center}         
	{
 \includegraphics[width=5.7cm,angle=0]{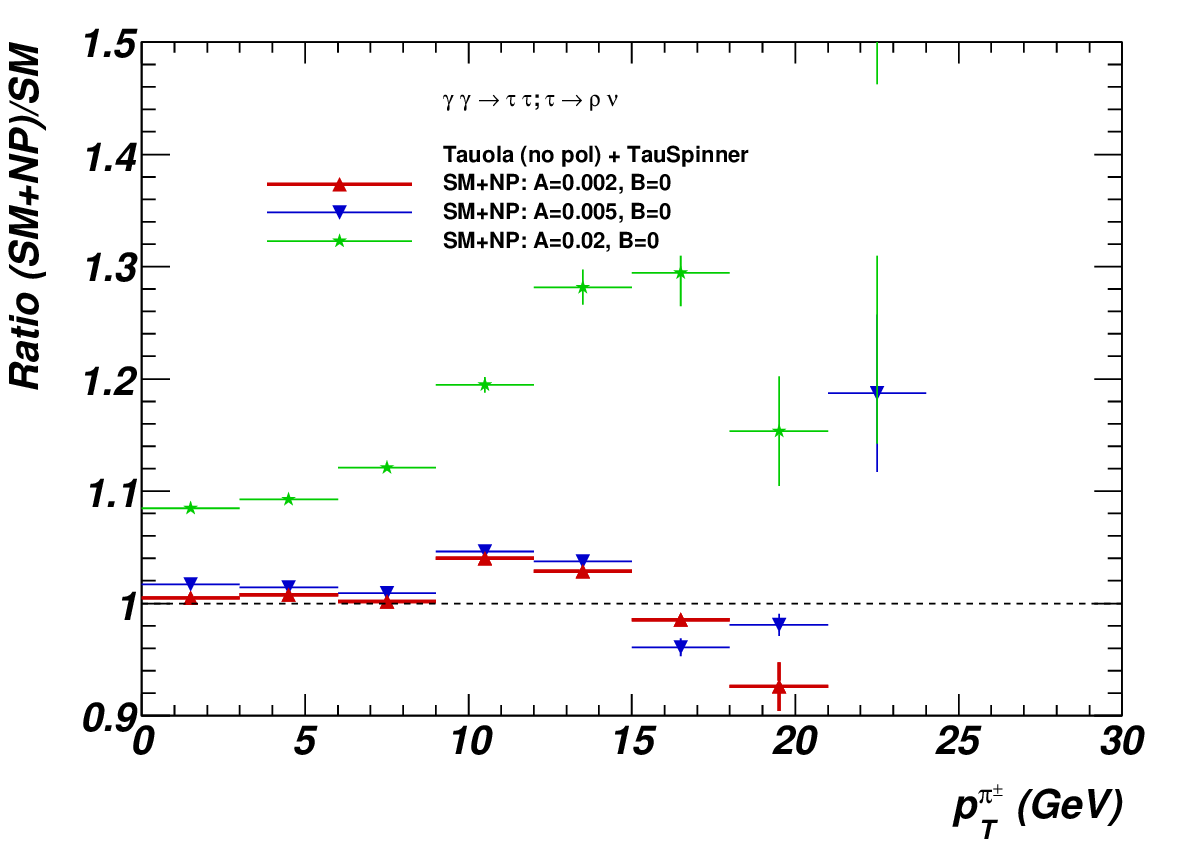}
 \includegraphics[width=5.7cm,angle=0]{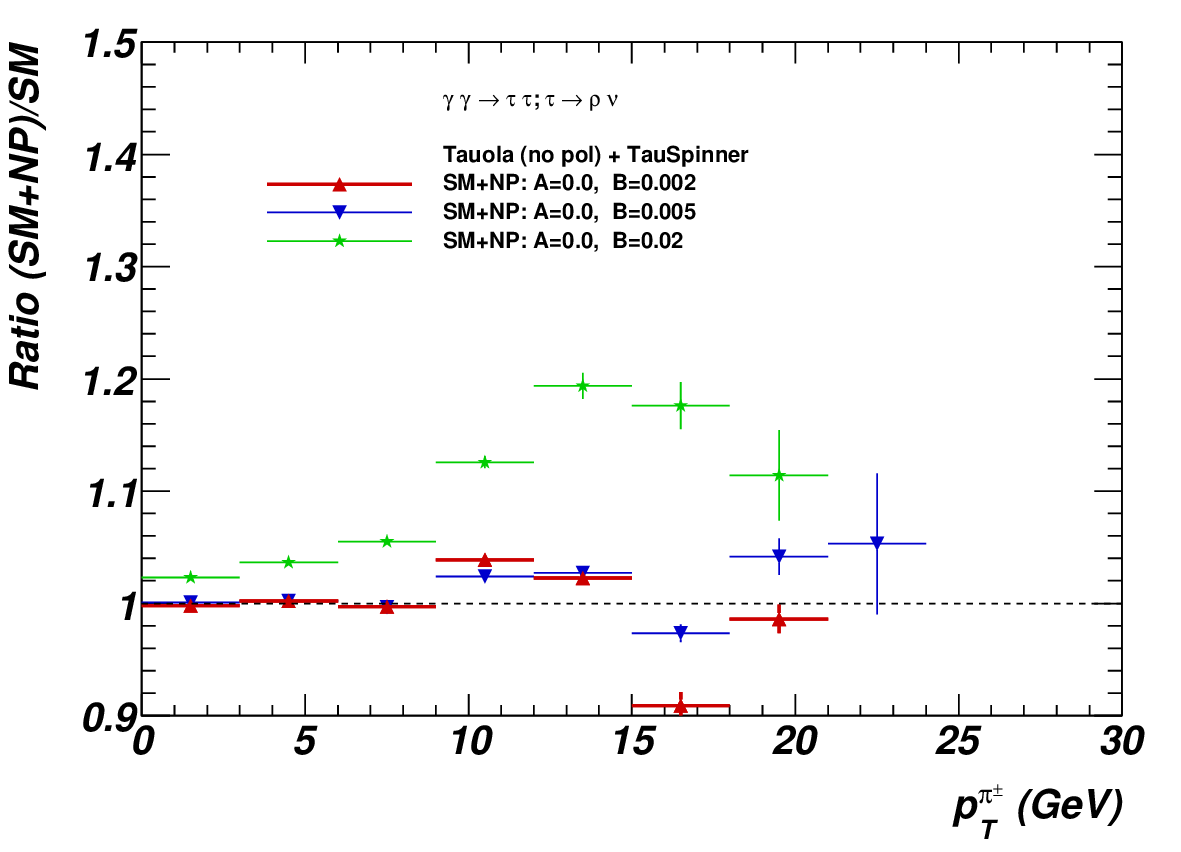} \\
 \includegraphics[width=5.7cm,angle=0]{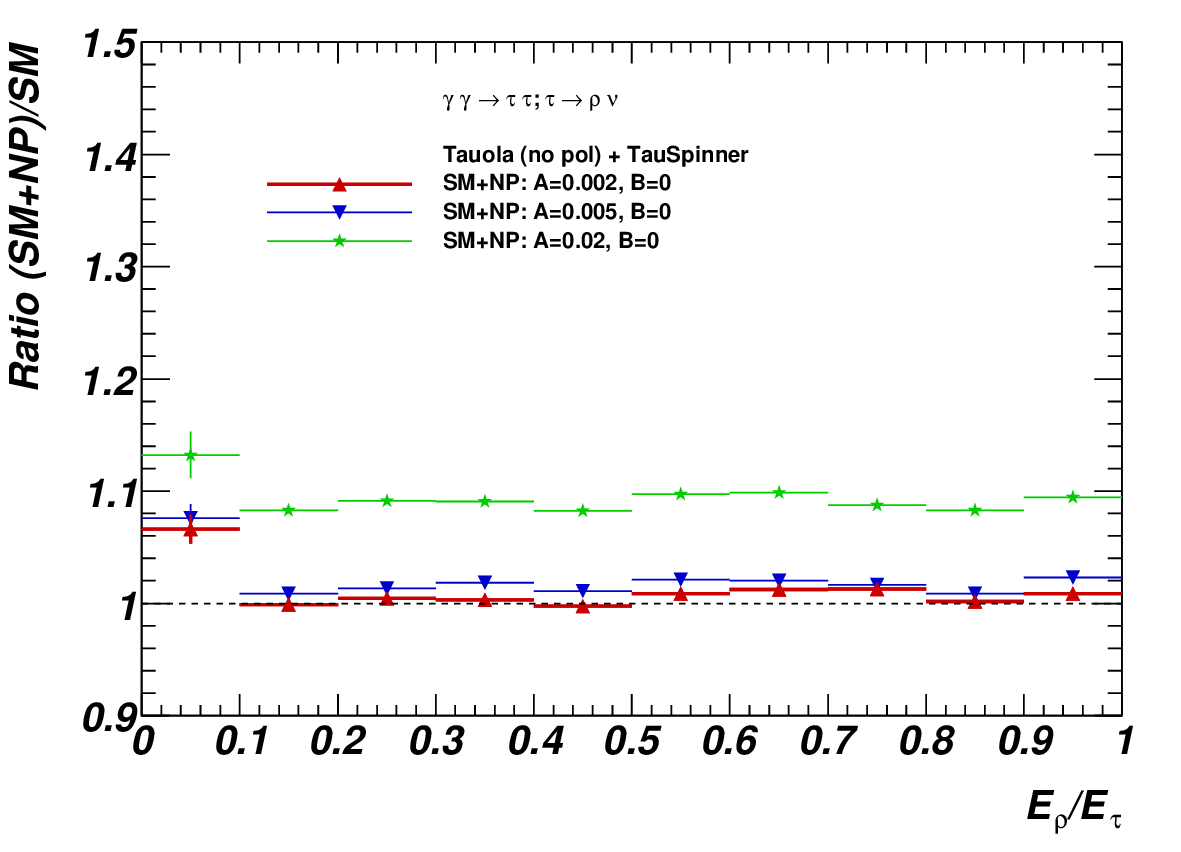}
 \includegraphics[width=5.7cm,angle=0]{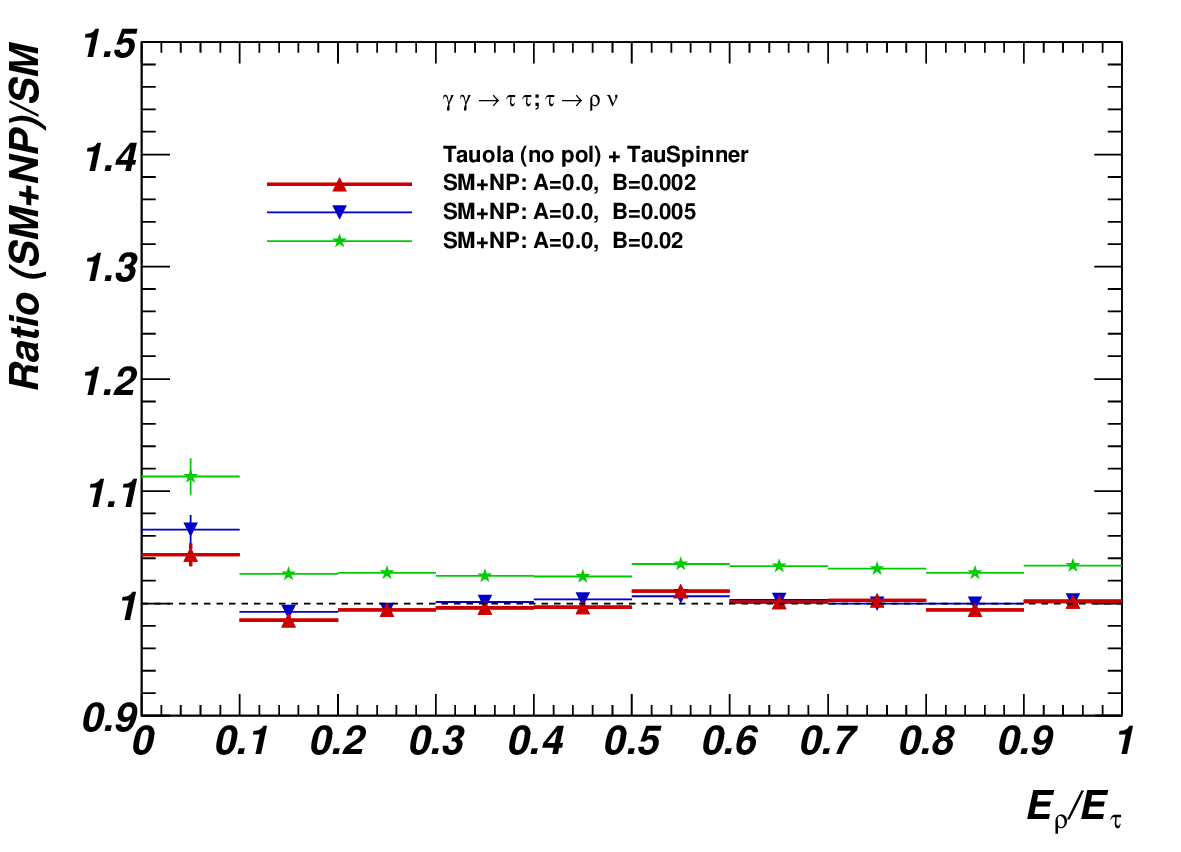} \\
 \includegraphics[width=5.7cm,angle=0]{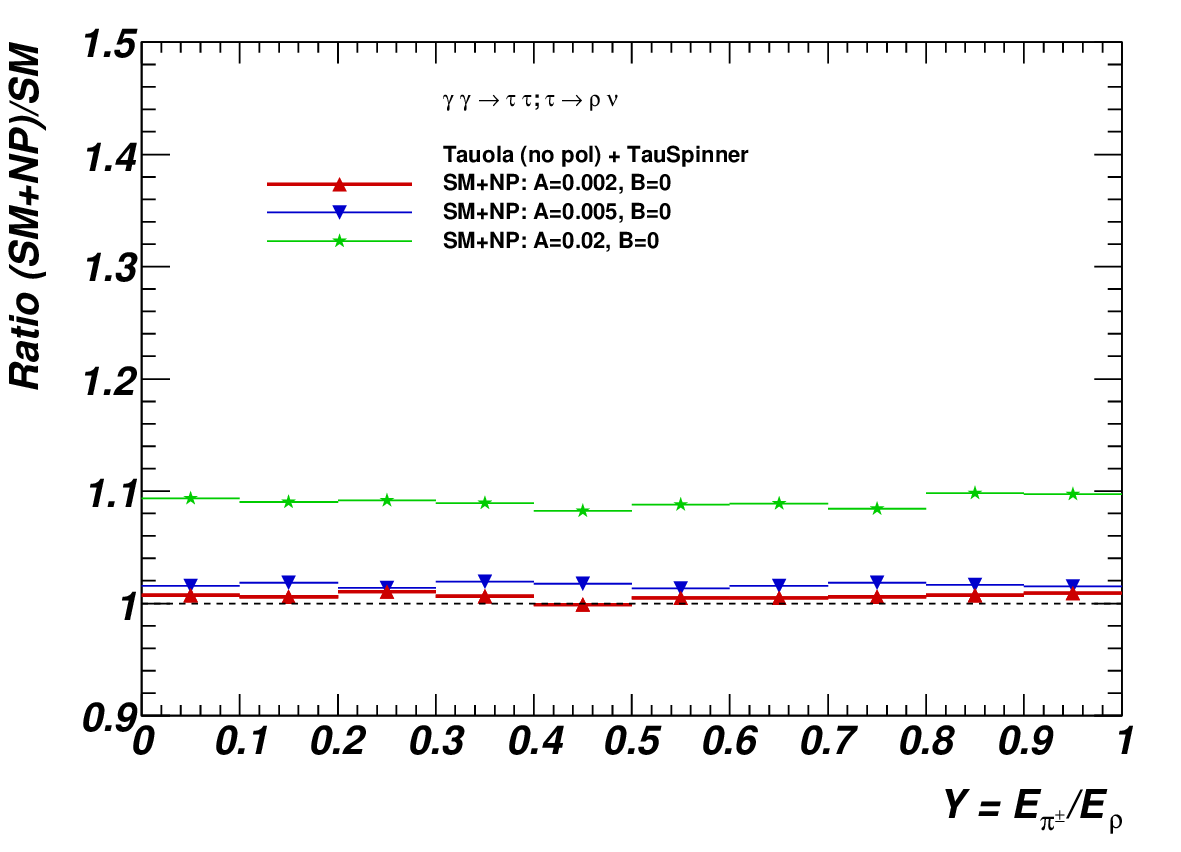}
 \includegraphics[width=5.7cm,angle=0]{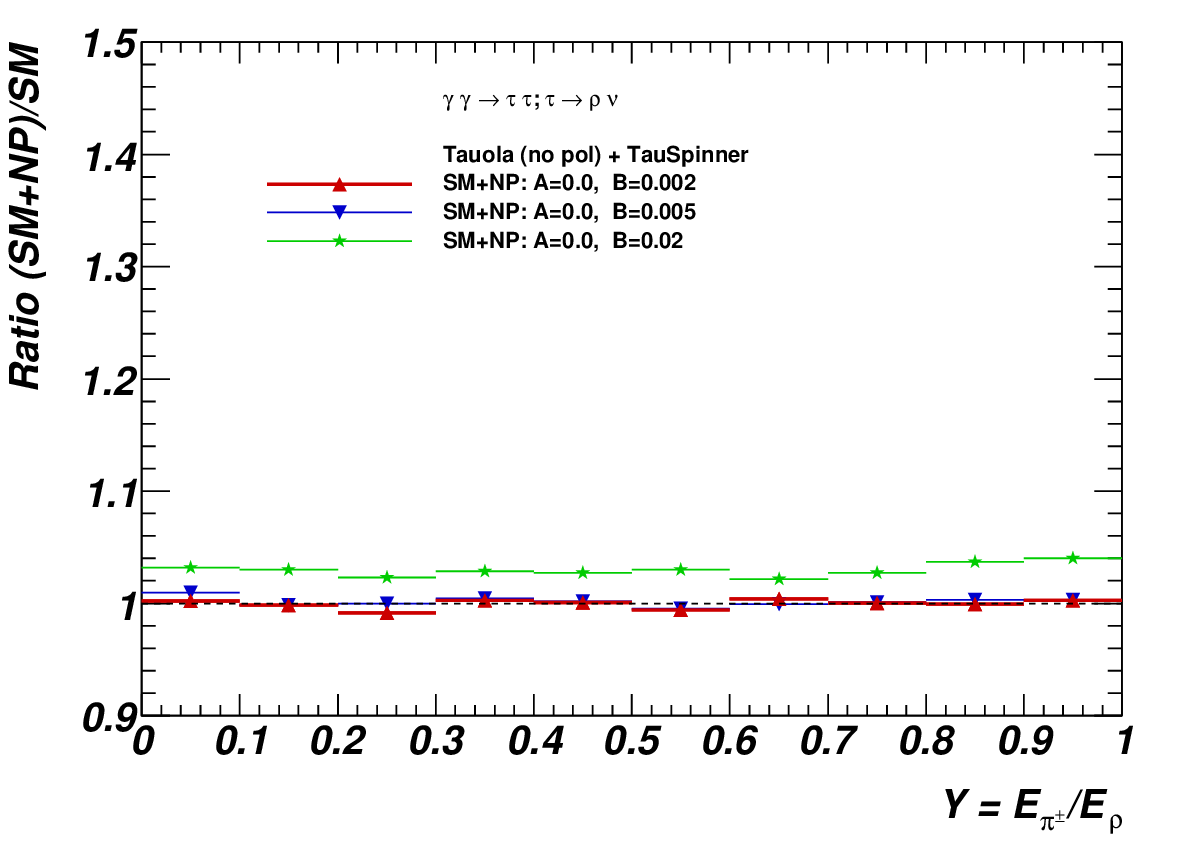} \\
 \includegraphics[width=5.7cm,angle=0]{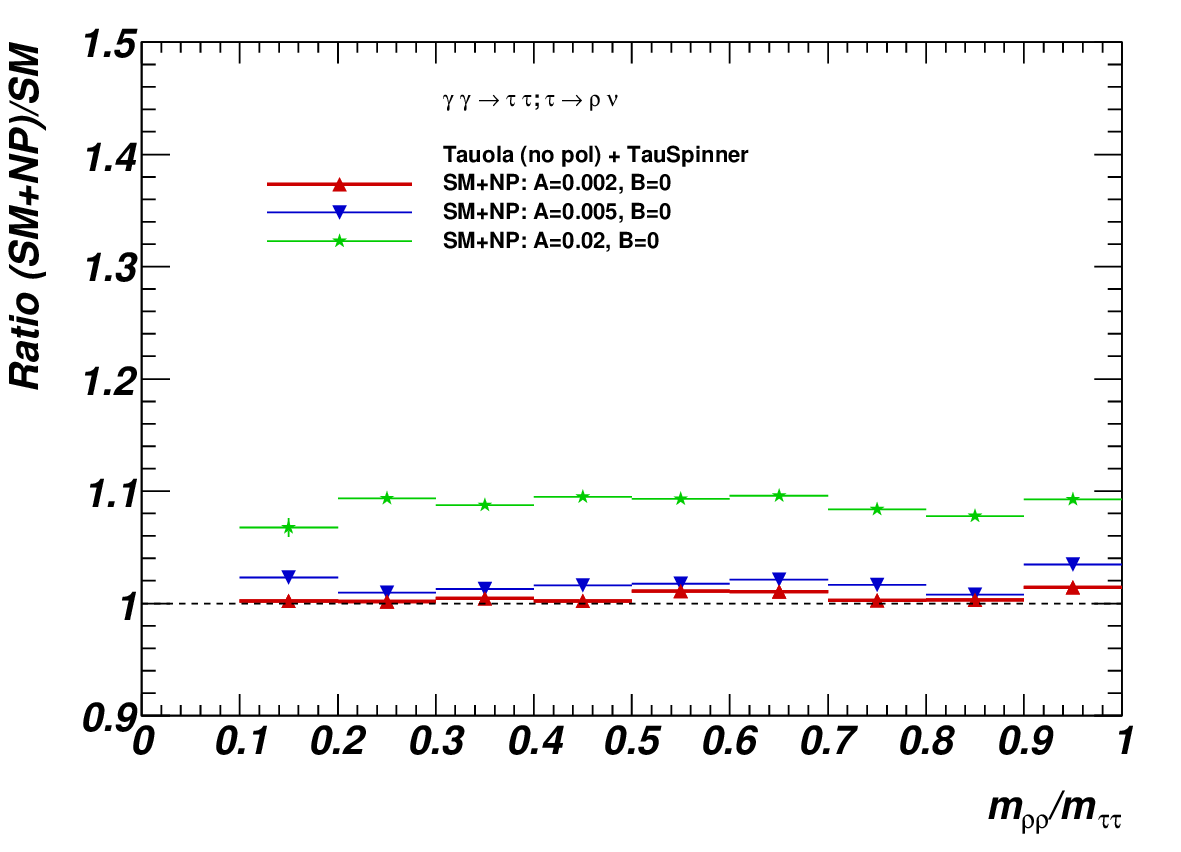}
 \includegraphics[width=5.7cm,angle=0]{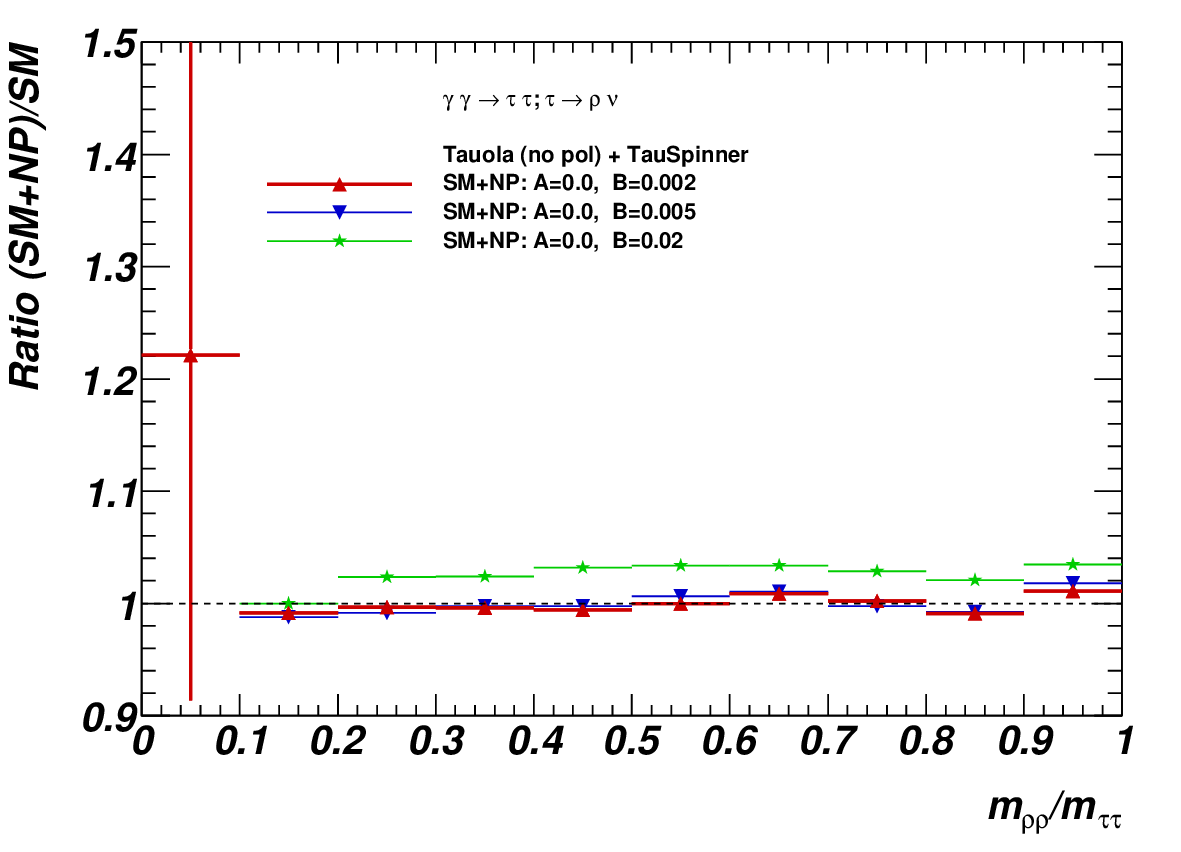}
}
\end{center}
  \caption{Kinematical distributions in the $\gamma \gamma \to \tau \tau$ process for $\tau$-lepton decays 
	$\tau^\pm \to \rho^\pm \nu_\tau \to \pi^\pm \pi^0 \nu_\tau$.  Compared are the calculations in SM and SM+NP
with the sets: \ $A = 0.002, \, 0.005, \, 0.02$, $B = 0$ (left column) and $A = 0.0$, 
$B = 0.002, \, 0.005, \, 0.02$ (right column).	 
The figure is taken from Ref.~\cite{Korchin:2025vzx}.}	
 \label{Fig:kinem_BSM_rhorho} 
\end{figure}

Fig.~\ref{Fig:kinem_BSM_rhorho} shows the effect of SM+NP including non-zero dipole moments and spin correlations.
Once integrated over the full phase-space, the impact from SM+NP extension is mostly due to the 
change of the cross section. 
Still changes in the shape of $p_T^{\pi}$ distribution are observed, 
with the ratio (SM+NP)/SM increasing with increasing $p_T^{\pi}$. 
Some effects are also visible in the $E_{\rho}/E_{\tau}$, $E_{\pi^\pm}/E_{\rho}$ and 
$m_{\rho \rho}/m_{\tau\tau}$ distributions for $A=0.02$. However, without 
detector studies it is not  clear how useful these effects can be for measurements searching for effects of NP.


\subsection{\texorpdfstring{Spin-correlation effects in $q \bar{q} \to \tau \tau$ process}{}}
\label{subsec:q-qbar}

Results for the $q \bar q \to \tau \tau $ process are also based 
on the events generated by {\tt Pythia 8.3}~\cite{Bierlich:2022pfr} 
for $pp$ collisions at 13 TeV, and the $\tau \tau$ invariant-mass range $m_{\tau\tau}$= 65-150 GeV. 
The $\tau$ decays $\tau^\pm \rightarrow \rho^\pm \nu_\tau$ are modeled with the {\tt Tauola} decay 
library~\cite{Jadach:1993hs}, and correlations were added using weight calculated with 
the {\tt TauSpinner} program. 

In the chosen region of $\tau \tau$ invariant mass close to the $Z$-boson peak, there is practically no dependence on the form-factors $A(q^2)$ and $B(q^2)$, however this region can be suitable for searching 
for signatures of the weak form-factors  $X(q^2)$ and $Y(q^2)$. 

The polarization and spin-correlation effects are illustrated for a few variables 
 calculated from the momenta of $\tau$-decay products:  \\
(i) ratios $E_{\rho}/E_{\tau}$ and $E_{\pi^\pm}/E_{\rho}$, which are sensitive to 
longitudinal polarization, \\
(ii) ratio of invariant masses of $\rho^+\rho^-$ system and $\tau^+\tau^-$ system, 
$m_{\rho\rho}/m_{\tau\tau}$, which is sensitive to longitudinal spin correlations, \\
 (iii) two acoplanarity angles $\Psi$ and $\phi^*$, which are sensitive to transverse-transverse 
 and transverse-normal spin correlations.

The definition of the $\Psi$ angle is chosen following the LEP publications~\cite{ALEPH:1997wux, DELPHI:1997ssw}.
The direction of the initial quark (beam) is assumed to be along the z-axis in the laboratory frame.
The four-momenta of $\tau$-decay products and beam momentum are boosted to 
the rest frame of $\rho^+ \rho^-$ system. In this system, the vectors of momenta are 
rotated around z- and y-axes such that $\pi^-$ is along the z-axis, and
$\pi^+$ lies in the xz-plane. Then $\Psi$ is the angle between the beam direction and $\pi^-$ direction.

The $\phi^*$ angle was proposed for the Higgs $CP$ studies in~\cite{Desch:2003rw} and
was used in the measurements at the LHC~\cite{ATLAS:2022akr, CMS:2021sdq}. 
It is the angle between the planes spanned on $\pi^{\pm}, \pi^0$ momenta for each $\tau^\pm$ decay, calculated in the rest frame of the $\rho^+ \rho^-$ system.
In this case, to preserve sensitivity to the transverse spin correlations, the
sample has to be split depending on the angle $\alpha_\rho$ between the beam axis and
plane spanned over $\pi^{\pm}, \pi^0$ from one decaying $\tau$ lepton, calculated in the laboratory frame
(splitting into regions: $\alpha_\rho <\pi/4$ and  $\alpha_\rho >\pi/4$). 

\begin{figure}[hbt!]
  \begin{center}                               
{
   \includegraphics[width=5.7cm,angle=0]{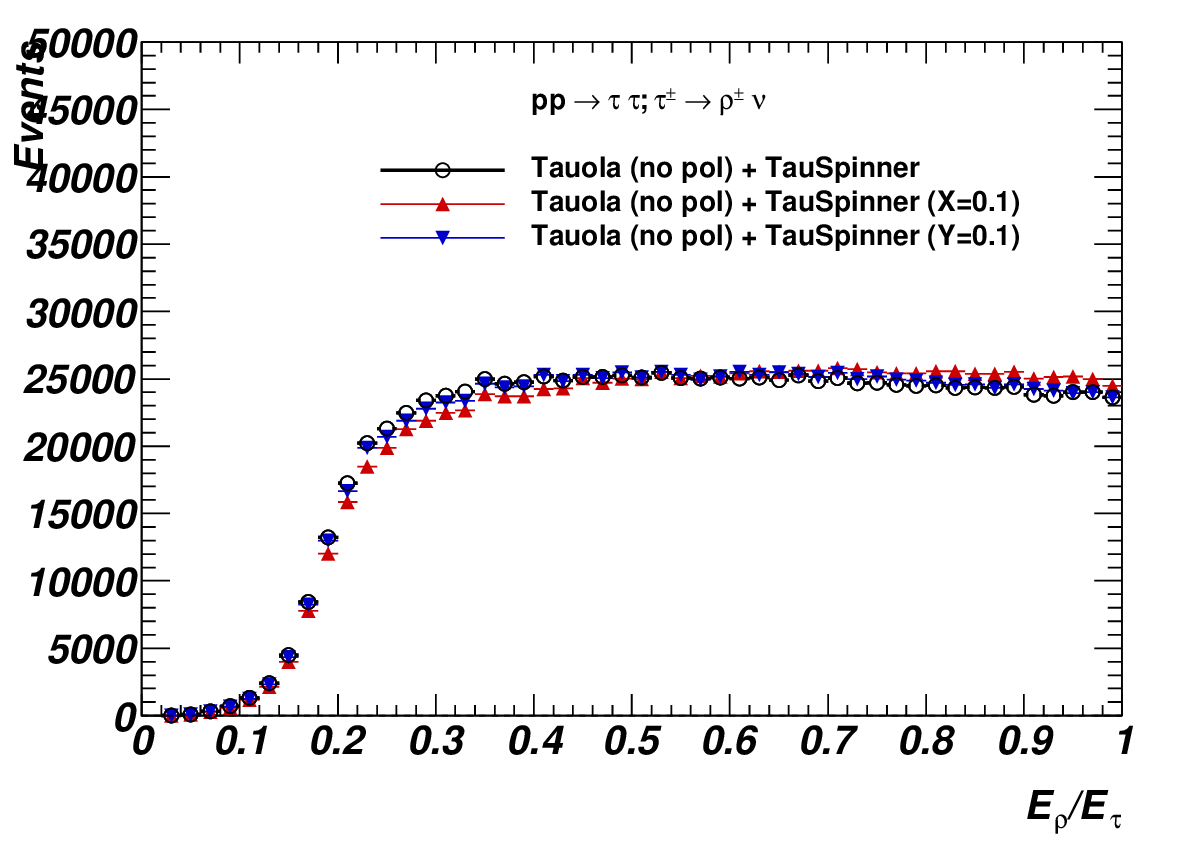}
   \includegraphics[width=5.7cm,angle=0]{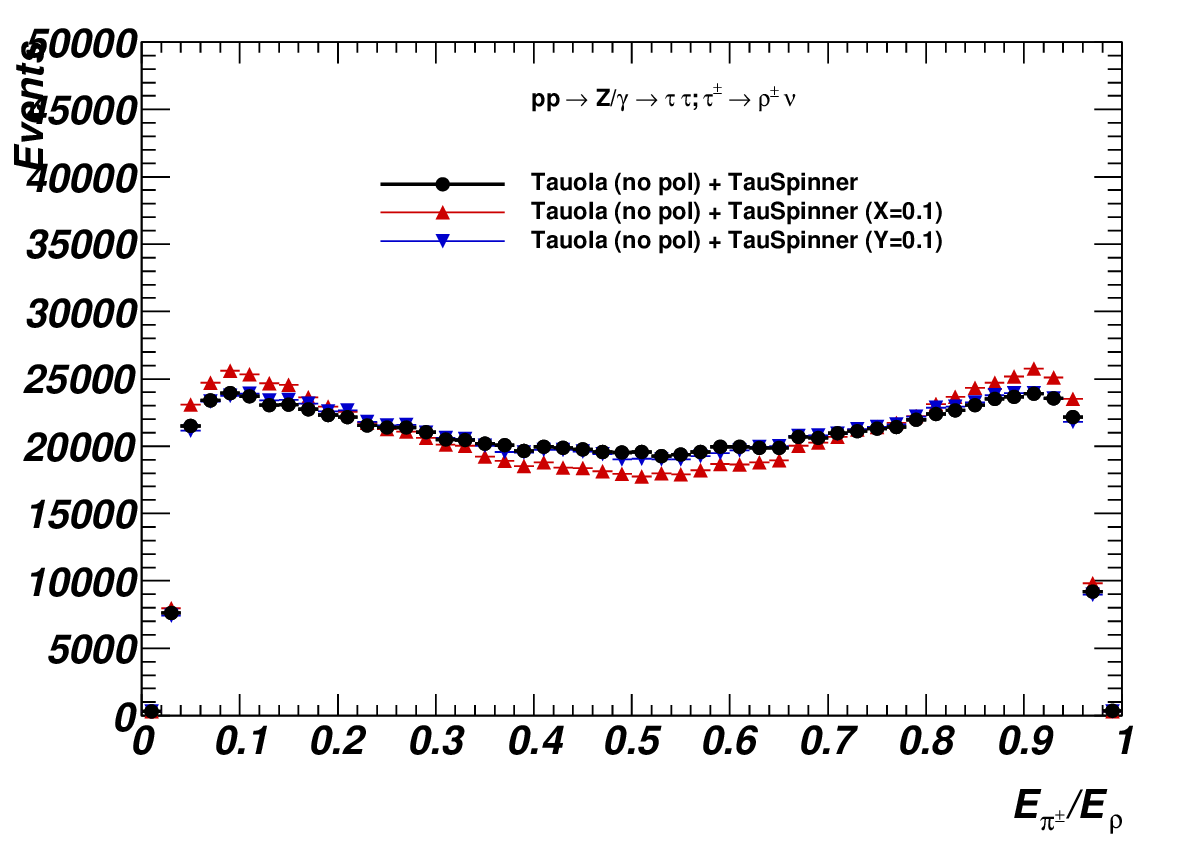}
   \includegraphics[width=5.7cm,angle=0]{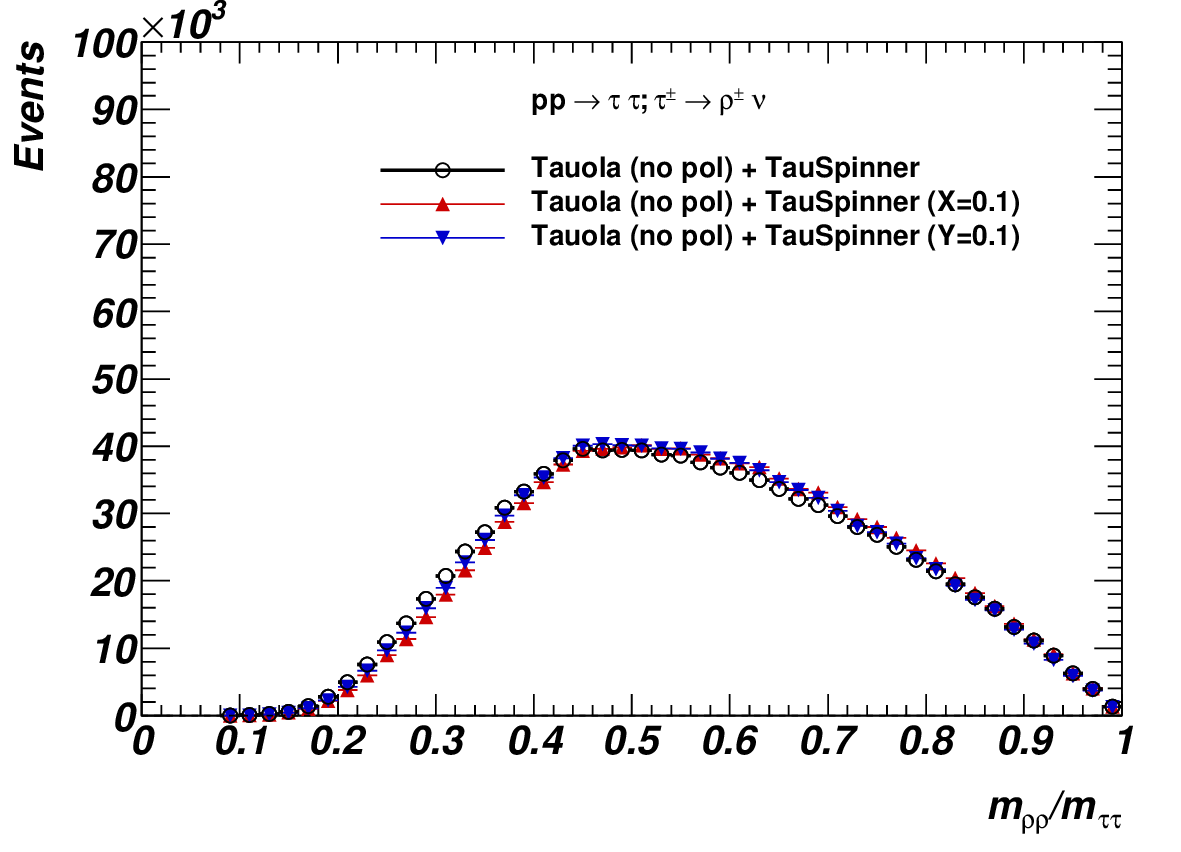} \\
   \includegraphics[width=5.7cm,angle=0]{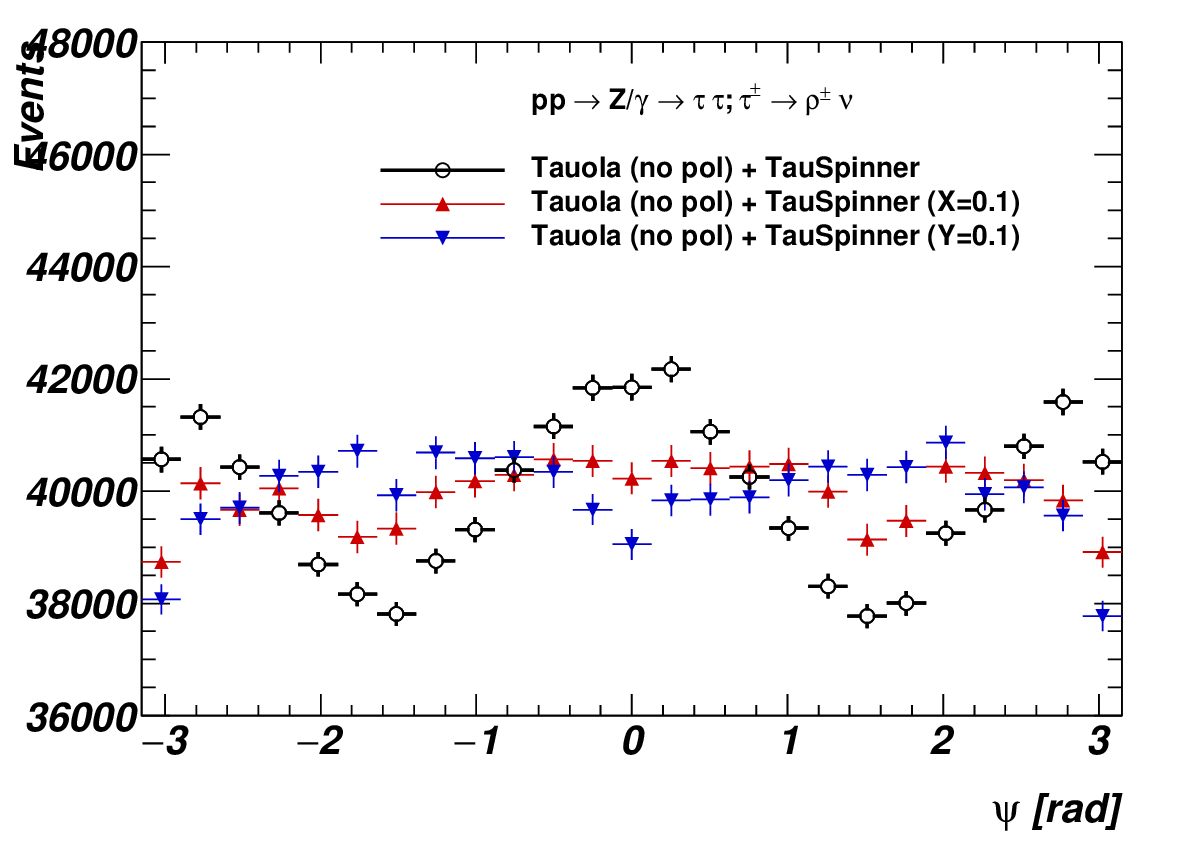}
   \includegraphics[width=5.7cm,angle=0]{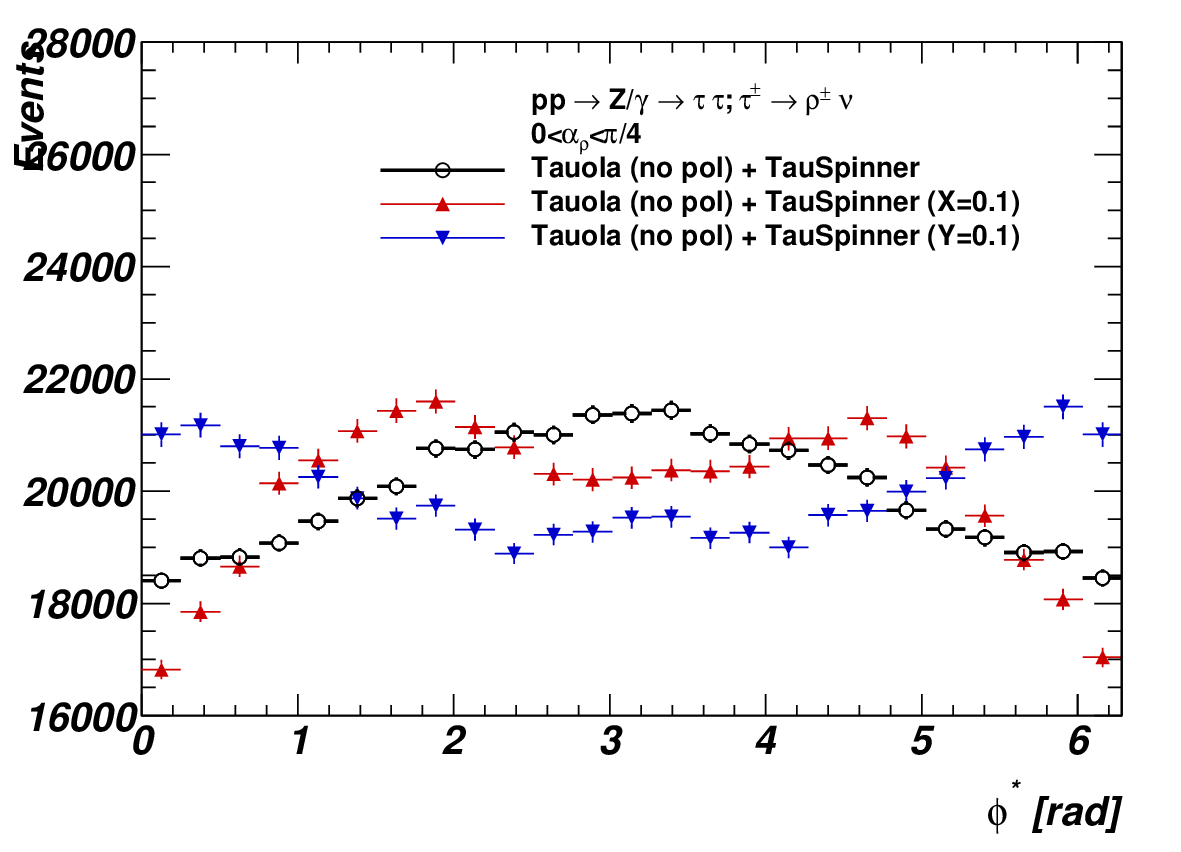}
}
\end{center}
\caption{Distribution of several observables (see the text) for the $q \bar{q} \to \tau \tau$ process.
Compared are the SM predictions (black open circles) for $X=Y=0$ and NP ones (red and blue triangles) for  
$X=0.1$ or $Y=0.1$.  Both $\tau$ leptons decay into the  
$\tau^\pm \to \rho^{\pm} \nu_\tau \to \pi^\pm \pi^0 \nu_\tau$ channels. The figure is taken 
from Ref.~\cite{Korchin:2025sui}.}
 \label{Fig:Kinem_rhorho_XY} 
\end{figure}

Distribution of spin-correlation sensitive observables in the SM with $X=Y=0$ and in the 
NP models with non-zero weak dipole moments is shown in Fig.~\ref{Fig:Kinem_rhorho_XY}.
Distributions of $E_{\rho}/E_{\tau}$,  $E_{\pi^\pm}/E_{\rho}$ and  $m_{\rho\rho}/m_{\tau\tau}$ 
show some effect of changed component $r_{tz}$ for $X=0.1$.  

Both  $\Psi$ and  $\phi^*$ distributions have cosine-like shape, once the sample is split into the 
two sub-samples depending on the sign of the product $y_{\tau^+} \cdot y_{\tau^-}$,  
where $y_{\tau^\pm}= (E_{\pi^\pm}-E_{\pi^0})/(E_{\pi^\pm}+E_{\pi^0})$.
If $y_{\tau^+} \cdot y_{\tau^-} < 0$, the angles are shifted as follows:
$\Psi \rightarrow \Psi -\pi/2$ and $\phi^* \rightarrow \phi^* -\pi$. 
This shift is included into definition of $\Psi$ and $\phi^*$, and then the final adjustment is made
to respect periodicity for $-\pi \leq \Psi \leq \pi $ and  $0 \leq \phi^* \leq 2 \pi$.
The periodical characteristic of the $\Psi$ and $\phi^*$ distributions
arises due to non-zero components $r_{xx}$, $r_{yy}$ of the spin-correlation matrix.
At the same time the phase of the distributions is sensitive to non-diagonal components
$r_{yx}$, $r_{xy}$. 

The dipole moments influence $\Psi$ and $\phi^*$ distributions either by flattening 
cosine-like pattern in the $\Psi$ distribution, or by changing the phase in the $\phi^*$ distribution. 
These effects are due to the changes in $r_{yy}$ for $X=0.1$, and in $r_{xy}$ for $Y=0.1$. 
However, the interplay between changes in the spin-correlation elements and distributions of  
acoplanarity angles seems nontrivial. 


\section{Summary}
\label{sec:Outlook}

In this article, the approach to implementing effects of electromagnetic and weak dipole moments (form-factors) 
in the $\gamma \gamma \to \tau \tau$ and $q \bar{q} \to \tau \tau$ processes is reviewed.
This topic is relevant in view of the experiments on the $\tau$-pair production in $pp$ and PbPb collisions
at the LHC.  The approach is based on the inclusion of spins of the final $\tau$ leptons in the parton-level
$2\to 2$ processes.   

The components of spin-correlation matrices in $pp \to \tau \tau $ production with subsequent $\tau$ decays
are studied in order to determine which of them are sensitive to the dipole moments, and can be useful 
for constructing kinematical observables.     
 
Several semi-realistic observables were studied, in particular, those built on momenta of the 
pions from the $\tau$-lepton decays $\tau^+ \to \rho^+ \bar{\nu}_\tau \to \pi^+ \pi^0 \bar{\nu}_\tau$,   
$\tau^- \to \rho^- \nu_\tau \to \pi^- \pi^0 \nu_\tau$. These decay channels of the $\tau$ lepton 
are selected due to sensitivity to the transverse components of the spin-correlation matrix. 
The consideration is focused on the transverse spin correlations as they are not  
often discussed in the literature, in particular, in $pp$ collisions.

This analysis is performed for the $\gamma \gamma \to \tau \tau$ and $q \bar{q} \to \tau \tau$
parton processes in the $pp$ collisions at the LHC energies. 
In the $q \bar{q} \to \tau \tau$  process near the $Z$-boson peak, the dipole moments are included on top of  
the Standard Model amplitudes supplemented with electroweak corrections in the
framework of the Improved Born Approximation. 
These ingredients were installed into the {\tt TauSpinner} event reweighting algorithm. 

It is shown that exploring spin effects in the final $\tau$ pair can enhance the sensitivity 
of some observables in high-energy experiments to New Physics signatures, 
in particular, to the anomalous magnetic and electric dipole moments (form-factors) 
of the $\tau$ lepton.   
\vspace{0.8cm}

 \noindent
{\bf Acknowledgments} \vspace{0.2cm}

The author would like to thank the organizers of the Conference ``Matter to the Deepest'' 2025 
in Katowice, Poland for the invitation to present the talk and hospitality. 
The author is grateful to Elzbieta~Richter-Was and Zbigniew~Was for fruitful collaboration on 
the topics discussed in this talk.  This project was supported in part by funds of the 
National Science Centre (NCN), Poland, grant No. 2023/50/A/ST2/00224 and
of the COPIN-IN2P3 collaboration with LAPP-Annecy, France.

 \bibliographystyle{utphys_spires}
\bibliography{Mttd2025_spin_tau}

\end{document}